\documentclass[journal=jctcce,manuscript=letter,layout=twocolumn]{achemso}
\setkeys{acs}{usetitle = true}

\usepackage{amssymb}
\usepackage{amsmath}
\usepackage{amsfonts}
\usepackage{graphicx}

\newcommand{\fancy}{\mathcal}

\newcommand{\FL}{\fancy{L}}

\newcommand{\Dp}[1]{\partial_{#1}}

\newcommand{\tr}{\text{Tr}}
\newcommand{\Tr}[1]{\tr\big[ #1 \big]}
\newcommand{\half}{\frac{1}{2}}

\newcommand{\around}{\sim\!\!}

\newcommand{\xc}{\text{xc}}

\newcommand{\Hrm}{\text{H}}
\newcommand{\xrm}{\text{x}}
\newcommand{\crm}{\text{c}}

\newcommand{\Ec}{E_{\crm}}

\newcommand{\dRPA}{\text{dRPA}}

\newcommand{\etal}{\emph{et al}}

\newcommand{\onlinecite}[1]{\hspace{-1 ex} \nocite{#1}\citenum{#1}}

\newcommand{\rcites}[1]{refs.\ [\onlinecite{#1}]}

\renewcommand{\vec}[1]{\mathbf{#1}}

\newcommand{\vr}{\vec{r}}
\newcommand{\vrp}{\vr^{\prime}}

\newcommand{\chih}{\hat{\chi}}
\newcommand{\wh}{\hat{w}}
\newcommand{\vh}{\hat{v}}
\newcommand{\uh}{\hat{u}}
\newcommand{\fh}{\hat{f}}
\newcommand{\fhx}{\fh_{\xrm}}

\newcommand{\xl}{\Theta_{\lambda}}

\newcommand{\disp}{\text{disp}}
\newcommand{\CP}{\text{CP}}
\newcommand{\ACFD}{\text{ACFD}}

\newcommand{\BtoA}{\text{$B \Leftrightarrow A$}}

\usepackage{color,xcolor}
\usepackage[normalem]{ulem}
\usepackage{soul}

\newcommand{\comment}[1]{}

\def\shortcomments{1}

\ifdefined\shortcomments

\fi

\ifdefined\nocomments

\fi

\title{Casimir Polder size consistency -- a constraint violated
by some dispersion theories}

\author{Tim Gould}
\email{t.gould@griffith.edu.au}
\affiliation{Qld Micro- and Nanotechnology Centre, Griffith
  University, Nathan, Qld 4111, Australia}
\author{Julien Toulouse}
\affiliation{Laboratoire de Chimie Th\'eorique, Universit\'e Pierre et
  Marie Curie, Sorbonne Universit\'es, CNRS, F-75005, Paris, France}
\author{J\'anos~G.~\'Angy\'an}
\affiliation{Laboratoire de Cristallographie, R\'esonance Magn\'etique et Mod\'elisations (CRM2, UMR CNRS 7036), Institut Jean Barriol, Universit\'e de Lorraine, F-54506 Vand{\oe}uvre-l\`es-Nancy, France}
\author{John F. Dobson}
\affiliation{Qld Micro- and Nanotechnology Centre, Griffith
  University, Nathan, Qld 4111, Australia}
\email{j.dobson@griffith.edu.au}

\begin{document}

\begin{abstract}
  A key goal in quantum chemistry methods, whether \emph{ab initio} or
  otherwise, is to achieve size consistency. In this manuscript we
  formulate the related idea of ``Casimir-Polder size consistency''
  that manifests in long-range dispersion energetics.
  We show that local approximations in time-dependent density
  functional theory dispersion energy calculations violate the
  consistency condition because of incorrect treatment of highly non-local
  ``xc kernel'' physics, by up to 10\% in our tests on closed-shell
  atoms.
\end{abstract}

\vspace{0.5cm}
Quantum chemical approaches and electronic structure theories more
generally aim to reproduce the key energetic physics of electrons
with the goal of calculating energies for systems of interest.
To a leading approximation two infinitely-separated quantum systems
should have an energy that is given by the sum of the energies of
the two components calculated separately -- a feature known as
size consistency.
Thus, quantum chemistry methods are generally expected to reproduce
this important property of quantum mechanics. Although its violation
is sometimes tolerated (see e.g. Nooijen \etal\cite{Nooijen2005})
for greater accuracy or lower cost,
it is nonetheless broadly accepted that size consistency is
an important goal in method development as it captures a fundamental
property of electronic systems.

The size consistency concept does not just apply at leading order,
however. As two systems $A$ and $B$ approach each other, additional
terms contribute to the energy, and these terms depend on properties
of the \emph{isolated} individual systems and the distance $D$
between them. As $D\to\infty$, the energy may thus be written as
\begin{align}
  E^{AB}(D) \sim& 
  E^A+E^B+U^{AB}(D).
  \label{eqn:EAB}
\end{align}
where the potential energy
\begin{align}
  U^{AB}(D) \sim&
  \sum_{p\geq 1} \frac{-C_{p}[\{\FL^A\},\{\FL^B\}]}{D^p}
  \label{eqn:UAB}
\end{align}
depends in some factorizable way \emph{only} on local properties
$\FL_p^X$ of the isolated systems $X=A,B$.
Thus, e.g. for systems with net local charges $Q^A$ and $Q^B$,
we have a leading term $U^{AB}(D)\to Q^AQ^B/D$ (i.e $C_1=-Q^AQ^B$).

Dipoles and higher multipoles yield similar expressions but with
larger exponents $p>1$ and thus decay more rapidly. These static
and multipolar contributions, including the static induction energy,
are present at the electrostatic level
and are properly included, via the Hartree energy, in all
size consistent quantum chemical approximations the authors
could think of.
Note that induction is sometimes considered to be a correlation
effect. Here we consider it to be an electostatic effect as it
\emph{is present at the self-consistent Hartree level},
unlike dispersion.

The leading beyond-electrostatic term is the attractive London
dispersion (van der Waals) potential $U^{AB}_{\disp}(D)=-C_6D^{-6}$,
which is also the dominant asymptotic term for finite neutral
systems without a permanent dipole or quadrupole. The coefficient,
\begin{align}
  C_{6,\CP}=&\int_0^{\infty}\frac{d\omega}{\pi}
  3\alpha^A(i\omega)\alpha^B(i\omega),
  \label{eqn:C6CP}
\end{align}
is obtained using an expression known as the Casimir-Polder
formula\cite{CP} that is in the general form of \eqref{eqn:UAB}.
Eq.~\eqref{eqn:C6CP} can also be obtained by calculating
$C_{6,\CP}^{AB}=-\lim_{D\to\infty}D^6U^{AB}_{\CP}(D)$ from
\begin{align}
  U^{AB}_{\CP}=&-\int_0^{\infty}\frac{d\omega}{2\pi}
  \Tr{ \chih^A\vh^{AB}\chih^B\vh^{BA} }\;,
  \label{eqn:UCP}
\end{align}
sometimes called the generalized Casimir-Polder
formula\cite{Kevorkyants2014}
which applies to more general geometries.
In this form it involves the anistropic
density-density imaginary-frequency linear response functions
$\chih^{A/B}\equiv\chih^{A/B}(i\omega)$ of the isolated
systems, and the Coulomb potential $\vh^{AB/BA}$ between them.
Here and henceforth, products
$\hat{G}\hat{H}=\int d\vr G(\vr_1,\vr)H(\vr,\vr_2)$
indicate convolutions over space variables and the trace
$\tr[\hat{G}]\equiv \int G(\vr,\vr)d\vr$ is similarly defined.

Here the local variable $\FL^X\equiv
\alpha^X(i\omega)=-\frac13\tr[(xx'+yy'+zz')\chih]$, from
\eqref{eqn:UAB}, is the spherically averaged imaginary-frequency
dipole polarizability of the system $X$
and depends only on properties of $X$ calculated in isolation.
Eq.~\eqref{eqn:C6CP} has proved to be exceedingly useful in
practical calculations of dispersion
forces\cite{Dobson2000-Novel,Dobson2001,Grimme2004,
Becke2005-XDM,Grimme2006,Tkatchenko2009,
Grimme2010,Tkatchenko2012,Toulouse2013-RSH,
Gould2016-C6Vol,Gould2016-C6,Gould2016-FI}, which have been
attracting much interest lately (see e.g.
\rcites{Grimme2011-Review,Dobson2012-JPCM,
Eshuis2012,Woods2016-Review} and references therein)
because of their increasingly recognised role in the behaviour
of multiple chemical and material science processes.

Alternatively, we can adopt a direct route to calculating dispersion
energies. We recognise that dispersion forces are a
purely correlation effect -- that is, they are absent in the
Hartree and exchange energy terms which capture all electrostatic
effects, at least for closed shell systems.
Thus, $U^{AB}_{\disp}(D)=\Ec^{AB}(D)-\Ec^A-\Ec^B\to-C_6/D^6$, giving
\begin{align}
  C_{6,\Ec}=&-\lim_{D\to\infty} D^6\big[ \Ec^{AB}(D) - \Ec^A-\Ec^B \big]
  \label{eqn:C6Ec}
\end{align}
where we calculate $\Ec^{AB}$ for the combined system $AB$ separated at
distance $D$. Thus, any method that can calculate correlation energies
can be used to determine $C_6$ coefficients.

\begin{figure}
  \includegraphics[width=0.9\linewidth]{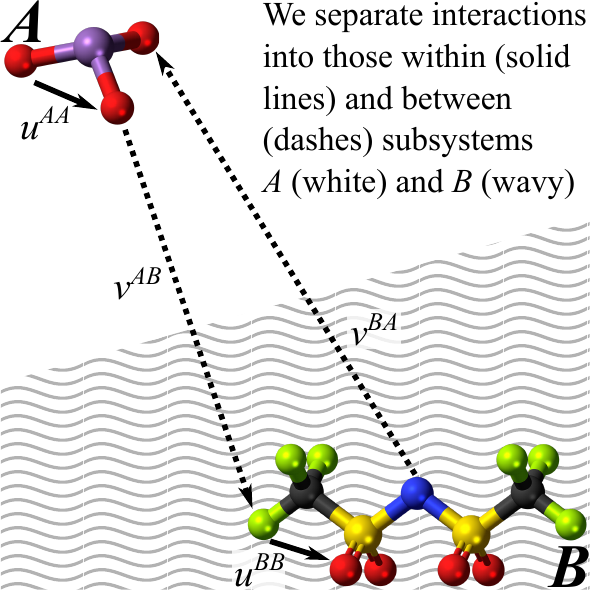}
  \caption{Interactions occur within ($u^{AA}$, $u^{BB}$)
    and between ($v^{AB}$, $v^{BA}$) systems $A$ and $B$.
    Switching off $v^{AB}/v^{BA}$ isolates the two systems.
  \label{fig:All}}
\end{figure}

This work now proceeds to formulate the idea of size consistency
of dispersion forces, called ``Casimir-Polder size
consistency''\cite{Dobson2012-Chapter}. Then, it will show how
time-dependent density functional theory\cite{RungeGross}
approximations can violate Casimir-Polder size consistency. Next,
it will give some examples illustrating the magnitude of the effect.
Finally, some conclusions will be drawn and impact discussed.

Let us first define Casimir-Polder size consistency.
Equations \eqref{eqn:C6CP} and \eqref{eqn:C6Ec} are obtainable
from first principles and thus should give the same
result, i.e. coefficients obtained from the Casimir-Polder formula
should be the same as those obtained from direct energy calculations.
Thus any theory for which \eqref{eqn:C6CP} equals \eqref{eqn:C6Ec}
is Casimir-Polder size consistent. Any approximation
where they are \emph{different} is not Casimir-Polder size consistent
and violates a fundamental property of well-separated systems.

We shall now proceed to show that, in time-dependent density
functional theory (TDDFT) calculations of dispersion energies
with a local exchange kernel,
the two approaches give different results, and thus such theories are
not Casimir-Polder size consistent.
Furthermore, other high-level quantum chemical approaches based
on screened response formalisms are also unlikely to be
Casimir-Polder size consistent. Such approaches are attracting interest%
\cite{PGG,Gould2012-RXH,Hellgren2013,Olsen2013-rALDA,%
Hesselmann2013-fxc,Jauho2015-rALDA,Dixit2016,Mussard2016}
because of their seamless inclusion of correlation physics, ability
to deal with metals and gapped systems, and moderate cost.
Any inconsistencies highlight a formal weakness of such approaches.

TDDFT offers two routes to dispersion energies.
Firstly, it can be used to calculate dipole polarizabilities for
use in the Casimir-Polder formula [Eq.~\eqref{eqn:C6CP}],
or the density-density linear response functions for \eqref{eqn:UCP}.
Secondly, it can be used to obtain correlation energies by using the
adiabatic connection formula\cite{ACFD} and fluctuation dissipation
theorem (ACFD). Energies thus obtained include dispersion forces
seamlessly\cite{Dobson1999,Furche2005,Eshuis2012}
[through Eq.~\eqref{eqn:C6Ec}],
making ACFD very useful for systems where dispersion competes with
other effects, in stark contrast to semi-local theories which do not
include \emph{any} long-range dispersion.

It thus serves as a go-to approach for attacking dispersion
calculations when beyond-empirical accuracy is required,
but when more advanced quantum mechanical methods are infeasible. For
example, TDDFT has been used to calculate $C_6$ coefficients of open
shell atoms and ions, giving good agreement with experiment and
more advanced methods\cite{Chu2004,Gould2016-C6}. A growing
number of researchers are using TDDFT and ACFD for increasingly
complex calculations\cite{Eshuis2012,Ren2012,Bjorkman2012,
Olsen2013-rALDA,Hesselmann2013-fxc,Jauho2015-rALDA,Dixit2016,Mussard2016}
that are not yet feasible in wavefunction methods.

\begin{table*}[h]
  \caption{$C_6$ coefficients calculated using ALDAx in
    Eqs.~\eqref{eqn:C6CP} and \eqref{eqn:C6Ec}. $\Delta C_6$
    quantifies the Casimir-Polder size consistency violation.
    \label{tab:Violation}}
  \begin{tabular*}{\linewidth}{@{}l@{\extracolsep{\fill}}|rrrrrrrr}
    \hline
    & He & Be & Ne & Mg & Ar & Ca & Zn & Kr \\
    \hline
    Eq.~\eqref{eqn:C6CP}~~~~
    & 1.39 & 260 & 5.62 & 695 & 63.7 & 2420 & 349 & 132 \\
    Eq.~\eqref{eqn:C6Ec}~~~~
    & 1.37 & 235 & 5.59 & 635 & 63.0 & 2170 & 332 & 130 \\
    $\Delta C_6$
    & 1.0\% & 9.5\% & 0.5\% & 8.6\% & 1.1\% & 10.5\% & 4.9\% & 1.3\% \\
    \hline
  \end{tabular*}
\end{table*}

The ACFD correlation energy,
\begin{align}
  E_{\crm}^{\ACFD}=&-\int_0^1d\lambda\int_0^{\infty}\frac{d\omega}{2\pi}
  \Tr{ (\chih_{\lambda}-\chih_0)\vh }
  \label{eqn:ACFD}
\end{align}
of an electronic system is given in terms of $\chih_0$, the linear
response of its density to changes in the effective potential $\hat{v}_s$;
and $\chih_{\lambda}$, the equivalent linear response to an external
potential at variable electron-electron interaction strength $\lambda$.
Notably, $\chih_1$ is the response of the real system
to the external potential,
and is the density-density linear response used in \eqref{eqn:UCP}.

The relationship between these response functions is
$\chih_{\lambda}=\chih_0
+ \chih_0[\lambda(\vh+\fh_{\xrm})+\fh_{\crm,\lambda}]\chih_{\lambda}$,
where all terms depend \textit{a priori} on $\vr$, $\vrp$ and $i\omega$, except for 
the Coulomb potential $\vh=1/|\vr-\vrp|$ which does not depend on $i\omega$.
$\fh_{\xrm}$ is the exchange kernel,%
\cite{Goerling1998-tdEXX,Hellgren2008,Hesselmann2010,%
Hellgren2010,Hellgren2012,Bleiziffer2012,Hellgren2013}
which is usually approximated. Finally, the correlation kernel
$\fh_{\crm,\lambda}$\cite{RungeGross} is defined similarly
to $\fh_{\xrm}$, but shall be assumed to be zero throughout
this manuscript.

We have so far kept the ACFD general. Let us now consider specifically
the $AB$ system, and introduce the ``locality'' assumption that
occurs in most TDDFT approximations, i.e. that the exchange kernel is
short-ranged in $|\vr-\vr'|$ and depends only on the properties of the
local system. We first partition space, as illustrated in
Figure~\ref{fig:All}, between systems $A$ and $B$ to
define $\uh=\sum_{X= A,B}(\vh^{XX} + \fhx^{XX})
\equiv \uh^{AA} + \uh^{BB}$, $\uh_{\Hrm}=\sum_{X=A,B}\vh^{XX}$
and $\wh=\vh^{AB}+\vh^{BA}$. Here
$\uh$ captures all intra-system interactions from both the Coulomb
$\vh^{XX}$ (corresponding to $\uh_{\Hrm}$)
and exchange kernel $\fhx^{XX}$ terms, where $\fhx^{XX}$ depends on
properties of system $X$ only. $\wh$ includes just the
long-ranged inter-system Coulomb interactions $\vh^{AB}/\vh^{BA}$ and
thus contains \emph{all dependencies} on $D$.
Then, we write the bare response $\chih_0=\chih_0^A + \chih_0^B$ as a
sum of subsystem responses $\chih_0^X$ calculated in isolation.

Note that for our present purposes we can now see that TDDFT
offers a conceptual advantage over wavefunction methods: both the
Casimir-Polder formula and the ACFD expression are well-defined for
any given kernel. Thus we can unequivocally talk about a subsystem
calculation of the polarizability, and a correlation energy calcuation
of the supersystem, at the same level of theory i.e. for a given
kernel approximation.

Now that the details of the different response functions and
interactions have been established, we shall next proceed to show
that coefficients calculated using
Eq.~\eqref{eqn:C6CP} [via \eqref{eqn:UCP}] are inconsistent
with coefficients obtained from \eqref{eqn:C6Ec} [via
\eqref{eqn:ACFD}] in a common class of approximations, which thus
lack Casimir-Polder size consistency.

With the assumptions described above, the TDDFT equation for the
response $\chih_{\lambda\gamma}$ of the combined system, with
intra-system interaction strength $\lambda$ and inter-system
interaction strength $\gamma$, is
\begin{align}
  \chih_{\lambda\gamma}=&\chih_{00} + \chih_{00}(\lambda \uh + \gamma \wh)
  \chih_{\lambda\gamma}\;,
  \label{eqn:chi}
\end{align}
where the bare response is $\chih_{00}\equiv \chih_0=\chih_{0}^A+\chih_{0}^B$.
Let us start with $\lambda=\gamma=0$ and first switch on the
intra-system interaction $\lambda$ while keeping $\gamma=0$
(equivalent to $D\to\infty$),
to obtain the isolated system response
$\chih_{\lambda 0}= \chih_{\lambda 0}^A+\chih_{\lambda0}^B
= \sum_X[1-\lambda \chih_{00}^X \hat{u}^X]^{-1}\chih_{00}^X$ from
\begin{align}
  \chih_{\lambda 0}=&\chih_{00} + \lambda \chih_{00} \uh \chih_{\lambda 0}\;.
  \label{eqn:chiXX}
\end{align}
Then we switch on the inter-system interaction $\gamma$ to obtain
\begin{align}
  \chih_{\lambda\gamma}=&
  \chih_{\lambda 0}+\gamma \chih_{\lambda 0} \wh \chih_{\lambda\gamma}\;.
  \label{eqn:chiXY}
\end{align}
It is readily verified that \eqref{eqn:chi} is reproduced by
substituting the solution of \eqref{eqn:chiXX} into \eqref{eqn:chiXY}.

Next we use \eqref{eqn:ACFD} to write
\begin{align*}
  U^{AB}=&\Ec^{AB}-\Ec^A-\Ec^B
  = -\int_0^{\infty}\frac{d\omega}{2\pi}\int_0^1d\lambda \xl\;,
\end{align*}
where the equivalence between $\gamma=0$ and $D\to\infty$ gives
$\xl=\tr[\chih_{\lambda\gamma}(\uh_{\Hrm}+\wh)
  - \chih_{\lambda0}\uh_{\Hrm}]_{\gamma=\lambda}$.
Iteration of \eqref{eqn:chiXY} to second order in $\gamma$
(since inter-system interactions $\wh$ are small)
then gives $\chih_{\lambda\gamma}\approx
\chih_{\lambda 0} + \gamma \chih_{\lambda 0}\wh \chih_{\lambda 0}
+\gamma^2 \chih_{\lambda 0} \wh \chih_{\lambda 0} \wh \chih_{\lambda 0}$,
and
\begin{align}
  \xl=&\Tr{ \lambda \chih_{\lambda 0} \wh \chih_{\lambda 0} \wh
  + \lambda^2 \chih_{\lambda 0} \wh \chih_{\lambda 0} \wh
  \chih_{\lambda 0}\uh_{\Hrm} }
  \label{eqn:xlambda}
\end{align}
to leading order. Here we dropped terms involving odd powers of
$\wh$ as these are exactly zero in the trace.

Let us now digress from the general formula to consider the direct
random-phase approximation (dRPA) which is the most popular,
albeit flawed, approach to the seamless calculation of molecular
and material properties using
ACFD\cite{Dobson2006,Harl2008,Gruneis2009,Dobson2009,
Lebegue2010,Eshuis2012,Bjorkman2012,Gould2016-Chapter}.
The dRPA consists of totally neglecting the exchange-correlation kernel
($f_{\xc}=0$), giving $\uh=\vh^{AA}+\vh^{BB}\equiv \uh_{\Hrm}$.
Taking the total derivative of \eqref{eqn:chiXX} gives
$\Dp{\lambda} \chih_{\lambda 0}=\chih_{\lambda 0}\uh_{\Hrm}\chih_{\lambda 0}$
so that \eqref{eqn:xlambda} becomes
$\xl^{\dRPA}=\half \Dp{\lambda}
\Tr{ \lambda^2 \chih_{\lambda 0}  \wh \chih_{\lambda 0} \wh } = 
\Dp{\lambda}(\lambda^2\tr[\chih_{\lambda0}^A\wh^{AB}\chih_{\lambda0}^B\wh^{BA}])$,
since $\wh$ cannot couple points in the same subsystem.
The occurrence of a perfect $\lambda$ derivative can be derived as
follows:
i) recognise that the explicit $O(\lambda)$ term can be expanded as
$\lambda\tr[\chih_{\lambda 0}^A\wh^{AB}\chih_{\lambda 0}^B\wh^{BA}
  +\BtoA]
=2\lambda\tr[\chih_{\lambda 0}^A\wh^{AB}\chih_{\lambda 0}^B\wh^{BA}]$;
ii) then use $\Dp{\lambda}\chih_{\lambda 0}^X
=\chih_{\lambda 0}^X\vh^{XX}\chih_{\lambda 0}^X$ in the explicit $O(\lambda^2)$
term to get $\lambda^2\tr[\chih_{\lambda 0}^A\wh^{AB}
  \chih_{\lambda 0}^B\vh^{BB}\chih_{\lambda 0}^B\wh^{BA} +\BtoA]
= \lambda^2\tr[\chih_{\lambda 0}^A\wh^{AB}
  [\Dp{\lambda}\chih_{\lambda 0}^B]\wh^{BA} + \BtoA]$,
which can be written using the cyclic properties of the trace as
$\lambda^2\Dp{\lambda}\tr[\chih_{\lambda 0}^A
\wh^{AB}\chih_{\lambda 0}^B\wh^{BA}]$;
iii) add the two terms to get $\xl^{\dRPA}=\Dp{\lambda}(\lambda^2
\tr[\chih_{\lambda 0}^A\wh^{AB}\chih_{\lambda 0}^B\wh^{BA}])$,
as desired.

Integrating over $\lambda$ then gives\cite{Dobson1994-vdW},
\begin{align}
  U^{AB}_{\dRPA}=&-\int_0^{\infty}\frac{d\omega}{2\pi}
  \Tr{ \chih_{\dRPA}^A \vh^{AB}\chih_{\dRPA}^B \vh^{BA} }\;,
  \label{eqn:UABdRPA}
\end{align}
to second order in $\wh$, which is Eq.~\eqref{eqn:UCP} calculated using
$\chih_{\dRPA}^X\equiv [1-\chih_{00}^X\vh^{XX}]^{-1}\chih_{00}^X$,
the dRPA response of the isolated system with full-strength
($\lambda=1$) internal Coulomb interaction $\vh^{XX}$.
Thus the energy calculated using the dRPA on the total system
[Eq.~\eqref{eqn:C6Ec}] is the same as that calculated using the
Casimir-Polder formula [Eq.~\eqref{eqn:UABdRPA} or \eqref{eqn:C6CP}]
with the dRPA response functions. The dRPA is Casimir-Polder size
consistent\cite{Dobson2012-Chapter}.

However, the dRPA is crude and relies on a cancellation of short-range
errors\cite{Furche2005} for its successes. Thus, work is ongoing
to improve on the dRPA by modelling the kernel%
\cite{PGG,Gould2012-RXH,Hellgren2013,Olsen2013-rALDA,%
Hesselmann2013-fxc,Jauho2015-rALDA,Dixit2016,Mussard2016}.
Let us now consider an exchange term $f_{\xrm}$ in our intra-system
interactions to get $\uh=\uh_{\Hrm}+\fh_{\xrm}$. Now,
$\Dp{\lambda}\chih_{\lambda 0}=\chih_{\lambda 0}(\uh_{\Hrm}+\fh_{\xrm})\chih_{\lambda 0}$
and we get
$\xl^{\xrm}=
\half\Dp{\lambda}\Tr{\lambda^2 \chih_{\lambda 0} \wh \chih_{\lambda 0} w}
  -\lambda^2 \Tr{\chih_{\lambda 0} \wh \chih_{\lambda 0}
    w \chih_{\lambda 0} \fh_{\xrm}}$
via a similar set of steps exhibited above for the dRPA.

Thus, in contrast to the dRPA, local TDDFT theories have an
additional term $\xl$ that cannot be written as a derivative.
After integration, the derivative term gives the expected
Casimir-Polder formula of Eq.~\eqref{eqn:C6CP} calculated with
the appropriate response
$\chih_{10}^{\xrm}=[1-\chih_{00}(\uh_{\Hrm}+\fh_{\xrm})]^{-1}\chih_{00}$
including the exchange kernel. The other term thus \emph{quantifies
the violation
of Casimir-Polder size consistency by the approximation}, which we
can express as
\begin{align}
  \Delta U^{AB}_{\xrm}
  &= \int_0^{\infty}\frac{d\omega}{2\pi}
  \int_0^1d\lambda \lambda^2
  \sum_{X=A,B} \nonumber\\&\times
  \Tr{\chih_{\lambda 0}^X \vh^{XY}
    \chih^Y_{\lambda 0} \fh^{YY}_{\xrm} \chih^Y_{\lambda 0} \vh^{YX} }
  \;,
  \label{eqn:DeltaU}
\end{align}
where $Y\neq X$ indicates the other system.

Eq.~\eqref{eqn:DeltaU} represents the key theoretical result of this
work, either directly or via its contribution
$\Delta C_6^{\xrm}=-\lim_{D\to\infty} D^6\Delta U_{\xrm}^{AB}(D)$ to the $C_6$
coefficient.
It illustrates that ACFD methods with beyond-Coulomb kernels acting
\emph{within} systems $A$ or $B$, but only Coulomb interactions 
acting \emph{between} systems $A$ and $B$ can give rise to a
difference in energies calculated using the Casimir-Polder formula
versus a full correlation energy calculation of the $AB$
system. Such approaches \emph{are not} Casimir-Polder size
consistent and thus violate a fundamental quantum mechanical
constraint.

We now investigate the magnitude of Eq.~\eqref{eqn:DeltaU} on a
selection of atomic systems using an adiabatic local density
approximation\cite{ALDA} for the exchange kernel only (ALDAx).
Thus, $f_{\xrm}(\vr,\vrp)\equiv\delta(\vr-\vrp)
f_{\xrm}^{\text{ALDA}}(n(\vr))$ where $f_{\xrm}^{\text{ALDA}}(n)$
is the second-order derivative of the exchange energy density of
the homogeneous electron gas with respect to the density $n$. This
kernel is chosen not for its
accuracy, but because it, like all semi-local kernels,
is obviously consistent with the assumptions we made about
$\uh^{XX}$ depending only on properties of system $X$,
and $\wh$ neglecting kernel terms entirely. 

It is worth noting that the local kernel used here produces a
divergent on-top correlation hole but a finite correlation energy.
Our general form \eqref{eqn:DeltaU} is not restricted to such
local kernels and can accommodate more accurate short-range
physics. The size-consistency issue is related to the long-range
physics, however, and is unlikely to be systematically improved
through better short-range physics.

Table~\ref{tab:Violation} reports $C_6$ values calculated (see
Gould and Bu\v{c}ko\cite{Gould2016-C6} for numerical details) using
Eqs.~\eqref{eqn:C6CP} and \eqref{eqn:C6Ec} within ALDAx, and shows
the difference as a percent. In some cases the difference between the
$C_6$ coefficients derived from the Casimir-Polder formula and the
energy of the system as a whole is substantial. For the highly
polarizable alkaline earth metals it can be as much as 10\% of the
total coefficient, a difference similar to the predicted accuracy of
TDDFT-derived coefficients\cite{Gould2016-C6}. By contrast, for noble
gases the difference is $\around1\%$, similar to numerical errors.

In conclusion, we have shown that local approximations to
TDDFT kernels violate a constraint we call ``Casimir-Polder
size consistency'', because the dispersion $C_6$ coefficient
calculated from properties of the two systems $A$ and $B$
[Eq.~\eqref{eqn:C6CP}] differs from that calculated, within
the same approximations, from the two systems studied together
[Eq.~\eqref{eqn:C6Ec}]. This result is inconsistent with ideas of
separability as manifested in Eqs.~\eqref{eqn:EAB} and \eqref{eqn:UAB}.
In the worst cases tested here, alkaline earth atoms, we find
significant deviations of $\around10\%$ using an exchange-only
adiabatic local density approximation. Worryingly, the deviation
seems to affect the most polarizable atoms the most, suggesting
its importance is \emph{amplified} in the very systems
where dispersion contributes most greatly to energetics.

Generalization of our results suggests that even sophisticated
``local'' correlation kernels (e.g. rALDA\cite{Olsen2013-rALDA})
cannot resolve the issue. We believe that similar problems will
manifest in some time-dependent generalized Kohn-Sham schemes
involving four-point kernels, although notably it was observed by
Szabo and Ostlund\cite{Szabo1977} that a variant of RPA
with a nonlocal Hartree-Fock exchange kernel is Casimir-Polder size
consistent (see also the discussion in
Ref.~\onlinecite{Toulouse2011-CCSD}). Work is ongoing to elucidate
more general cases, including important wavefunction methods.

Caution is thus advised when comparing long-range forces calculated
using polarizabilities, or via systems as a whole. Such approaches
include range-separated approaches.%
\cite{Toulouse2009,Toulouse2010,%
Zhu2010,Toulouse2011-CCSD,AngLiuTouJan-JCTC-11}
Guaranteeing Casimir-Polder size consistency should be a goal for
new kernel approximations\cite{Kevorkyants2014}.
Similarly, one might look for response models
that can \emph{reproduce} by construction quantum chemical theories
of supersystems and thus automatically avoid Casimir-Polder
size consistency issues.

\begin{acknowledgement}
The authors wish to acknowledge the important contribution of
our good friend and colleague, the late Prof. J\'anos \'Angy\'an.
This work would not exist without his efforts, and we are
saddened he could not see it completed.
T.G. received funding from a
Griffith University International Travel Fellowship.
\end{acknowledgement}


\providecommand{\latin}[1]{#1}
\makeatletter
\providecommand{\doi}
  {\begingroup\let\do\@makeother\dospecials
  \catcode`\{=1 \catcode`\}=2\doi@aux}
\providecommand{\doi@aux}[1]{\endgroup\texttt{#1}}
\makeatother
\providecommand*\mcitethebibliography{\thebibliography}
\csname @ifundefined\endcsname{endmcitethebibliography}
  {\let\endmcitethebibliography\endthebibliography}{}

\end{document}